# Compressive Direction Finding Based on Amplitude Comparison

Ruiming Yang, Yipeng Liu, Qun Wan and Wanlin Yang
Department of Electronic Engineering
University of Electronic Science and Technology of China
Chengdu, China
{ shan99, liuyipeng, wanqun, wlyang}@uestc.edu.cn

*Abstract*—This paper exploits recent developments in compressive sensing (CS) to efficiently perform the direction finding via amplitude comprarison. The new method is proposed based on unimodal characteristic of antenna pattern and sparse property of received data. Unlike the conventional methods based peak-searching and symmetric constraint, the sparse reconstruction algorithm requires less pulse and takes advantage of CS. Simulation results validate the performance of the proposed method is better than the conventional methods.

*Keywords - direction finding; amplitude comparison; beam scanning; sparse reconstruction; compressive sensing.*

## I. INTRODUCTION

With the development of radar technology and the complication of target background, more and more information which is not range but also angle need be known to target in order to track and orientate accurately. In most modern radar systems, the target direction of arrival is estimated by the monopulse technique [1], which in principle can work with just a single pulse.

Different from the direction-finding methods of monopulse radar, there is another method that works as follows: The beam of radar antenna scans to find the user; then the user responses; finally the radar measures the strength of the response signal, and finds the user's location to the radar by the modulation information of the pattern. As the radar antenna pattern has obvious peak features, so the user position relative to the radar can be determined directly using the estimated peak location method.

There are many ways to estimate the peak position. An efficient algorithm for estimating the peak position of a sampled function is the Hilbert Transform interpolation algorithm [2]. The algorithm is a computationally efficient algorithm for the peak detection and position estimation of a signal function. It is based on a signal interpolation technique which relies on the Hilbert Transform of the sampled signal. Besides, another method such as the multi-resolution method which is able to overcome the sampling period's influence on the peak position estimation accuracy, Fourier transform time shift invariant Methods and Sinc function interpolation method [3] can estimate the peak location too.

This paper re-examines the angle estimation problem and uses recent results in sparse approximation [4] and compressive sensing to provide a fundamentally different direction finding method. First we get a sparse representation of the received signal and then the user's location to radar is obtained by the sparse solution. Comparing with the traditional unimodal characteristic and symmetry constraints based maximum (SCBM) methods, the proposed one requires fewer pulses, is with the ability of compressed sampling, and achieves a much smaller estimation error than the traditional search method.

This paper is organized as follows. The Compressed sensing review is described in Section II. In section III we presented the measurements model. We introduce four direction finding methods in section IV: the traditional maximum method and symmetry constraints based maximum method, the match pursuit and basis pursuit methods which based on the compressive sensing. Section V presents simulation results that validate the formulation and demonstrate significant performance increase over traditional maximum methods. Conclusions are presented in section VI.

## II. COMPRESSED SENSING REVIEW

Sparsity widely exists in wireless signals [5]. Considering a signal **x** can be expanded in an orthogonal complete dictionary, with the representation as

$$\mathbf{x}_{N\times 1} = \mathbf{\Psi}_{N\times N}\mathbf{b}_{N\times 1}, \qquad (1)$$

when most elements of the vector **b** are zeros, the signal **x** is sparse. And when the number of nonzero elements of **b** is S (S << L < N), the signal is said to be *S*-sparse. Compressive Sensing (CS) provides an alternative to the well-known Shannon sampling theory. It is a framework performing non-adaptive measurement of the informative part of the signal directly on condition that the signal is sparse.

In CS, instead of measure the signal directly as Nyquist sampling, a random measurement matrix Φ is used to sample the signal. In matrix notation, the obtained random sample vector can be represented as

$$\mathbf{y}_{M\times 1} = \mathbf{\Phi}_{M\times N}\mathbf{x}_{N\times 1}, \qquad (2)$$

The measurement matrix should satisfy the restricted isometry property (RIP) which is a condition on matrices **Φ** which

This work was supported in part by the National Natural Science Foundation of China under grant 60772146, the National High Technology Research and Development Program of China (863 Program) under grant 2008AA12Z306 and in part by Science Foundation of Ministry of Education of China under grant 109139.*)*

provides a guarantee on the performance of $\mathbf{\Phi}$ in CS. It can be stated as:

$$(1-\delta_s)\|\mathbf{y}\|_2^2 \leq \|\mathbf{\Phi y}\|_2^2 \leq (1+\delta_s)\|\mathbf{y}\|_2^2, \quad (3)$$

for all *S*-sparse $\mathbf{y}$. The restricted isometry constant $\delta_s \in (0,1)$ is defined as the smallest constant for which this property holds for all s-sparse vectors $\mathbf{y}$.

There are three kinds of frequently used measurement matrices:
1) Non-Uniform Subsampling (NUS) or Random Subsampling matrices which are generated by choosing *M* separate rows uniformly at random from the unit matrix $\mathbf{I}_N$;
2) Matrices formed by sampling the i.i.d. entries $(\mathbf{\Phi})_{ij}$ from a white Gaussian distribution;
3) Matrices formed by sampling the i.i.d. entries $(\mathbf{\Phi})_{ij}$ from a symmetric Bernoulli distribution and the elements are $\pm 1/\sqrt{N}$ with probability 1/2 each.

When the RIP holds, a series of recovering algorithm can reconstruct the sparse signal [6]. One is greedy algorithm, such as matched pursuit (MP) [7], OMP [8]; another group is convex programming, such as basis pursuit (BP), LASSO and Dantzig Selector (DS) [9]. DS has almost the same performance as LASSO. Both of the convex programming and greedy algorithm have advantages and disadvantages when applied to different problem scenarios. A very extensive literature has been developed that covers various modifications of both algorithms so to emphasize their strengths and neutralize their flaws. A short assessment of their differences would be that convex programming algorithm has a more reconstruction accuracy while greedy algorithm has less computing complex. And in contrast to BP, LASSO has additional denoising performance advantage.

## III. MEASUREMENT MODEL

As described in section I, the radar antenna pattern has obvious peak features, so the user position relative to the radar can be determined directly using the estimated peak location method. Assume that the antenna pattern is $p(\theta)$. Without loss of generality, as shown in Fig. 1, let $p(\theta)$ be a Sinc function and represented as:

$$p(\theta) = \text{sinc}^2\left(\theta_b^{-1}\theta\right) \quad (1)$$

where $\theta_b$ is the half of the mainbeam width. In the *k*-th moment, the strength of the received signal can be represented as:

$$x_k = s_k + v_k \quad (2)$$

where $s_k = p(\theta_k)$; $\theta_k$ is the radar antenna scanning angle in the *k*-th moment; $v_k$ is the noise, which is in $\chi^2$ distribution, $k = 1, 2, \cdots, K$, $K$ is the number of received signal strength in the measurement period.

In (2), $s_k$ has only the information of the received signal strength along with the change of radar antenna's scanning angle. Thus $v_k$ is already normalized by the maximum received signal strength.

## IV. DIRECTION FINDING

### A. Maximum Method

Maximum method takes advantage of peak characteristic of the pattern $p(\theta)$. The user's relative azimuth to the radar is estimated by finding the location corresponding to the maximum element in the received signal's strength sequences $x_1, x_2, \cdots, x_K$, i.e.

$$\hat{\theta}_{\text{MAX}} = \theta_{\hat{k}}, \quad (3)$$

where

$$\hat{k} = \arg \max_{1 \leq k \leq K} (x_k), \quad (4)$$

### B. Symmetry Constraints Based Maximum Method

The symmetry constraints based maximum (SCBM) method uses both the unimodal characteristic and the symmetry. It estimates the user's azimuth relative to the radar by finding the best symmetrical location corresponding to the maximum element in the user's received signal strength sequences, i.e:

$$\hat{\theta}_{\text{SYM}} = \theta_{\hat{k}}, \quad (5)$$

where

$$\hat{k} = \arg \max_k \left( \frac{x_m \sum_{m=1}^{N} x_{k-m} x_{k+m}}{\sqrt{\sum_{m=1}^{N} x_{k-m}^2 \sum_{m=1}^{N} x_{k+m}^2}} \right), \quad (6)$$

N is the minimum element between $k-1$ and $K-k$, i.e.

$$N = \min(k-1, K-k), \quad (7)$$

### C. Match Pursuit

Searching over an extremely large dictionary for the best match is computationally unacceptable for practical applications. Mallat and Zhang proposed a greedy solution that is known from that time as Matching Pursuit. Matching pursuit is a type of numerical technique which involves finding the "best matching" projections of multidimensional data onto an over-complete dictionary. The basic idea is to represent a signal from Hilbert space as a weighted sum of functions (called atoms). By taking an extremely redundant dictionary we can look in it for functions that best match a signal. Finding a representation where most of the coefficients in the sum are close to 0 (sparse representation) is desirable for signal coding and compression.

Here we divide the beam width into $2L+1$ sections with the same length, i.e. $\theta_{-L}, \theta_{-L+1}, ..., \theta_k, ..., \theta_L$, where $\theta_k = \theta_b \cdot k / L$, $k = -L, -L+1, \cdots, 0, 1, \cdots, L$, and construct the Redundant dictionary as:

$$\mathbf{D} = \text{Toeplitz}(\mathbf{d_1}, \mathbf{d_2}) \quad (8)$$

where

$$\mathbf{d}_1 = [p(\theta_0) \quad p(\theta_1) \quad \cdots \quad p(\theta_L) \quad 0 \quad 0 \quad \cdots \quad 0]^T \quad (9)$$

$$\mathbf{d}_2 = [p(\theta_0) \quad p(\theta_{-1}) \quad \cdots \quad p(\theta_{-L}) \quad 0 \quad 0 \quad \cdots \quad 0] \quad (10)$$

where $p(\theta_k) = \text{sinc}^2(\theta_b^{-1}\theta_k)$, and the column vector $D_i (i=1,...,2L+1)$ of the redundant dictionary is named as atom.

$$\mathbf{x} = \mathbf{Ds} + \mathbf{v}, \quad (11)$$

where $\mathbf{x} = [x_1,...,x_K]^T$, $\mathbf{v} = [v_1,...,v_K]^T$. The user's azimuth to the radar can be determined by finding the non-zero element of vector $\mathbf{s}$, and the corresponding atom can be obtained by finding the maximum correlation of the atom and received signal $\mathbf{x}$, i.e.:

$$D_{opt} = \arg\max_i |\langle D_i, x \rangle|, \quad (12)$$

*D. Basis Pursuit*

To encourage sparsity, The $\ell_0$ optimization is optimal but non-convex and known to be NP-hard. In practice, a multitude of efficient algorithms have been proposed, which achieve high recovery rates. The $\ell_1$-minimization method is the most extensively studied recovery technique. In this approach, the non-convex $\ell_0$ norm is replaced by the convex $\ell_1$ norm. This approximation is known as Basis Pursuit (BP) which is a principle for decomposing a signal into an "optimal" superposition of dictionary elements, where optimal means having the smallest $\ell_1$ norm of coefficients among all such decompositions.

Here the sparse solution of (11) can be obtained by optimization method. It can be modeled as:

$$\min \|\mathbf{s}\|_1 + \sum_{i=1}^{K} e_i \quad s.t. \quad e_i \geq 0, \quad s_i \geq 0, \quad (13)$$

where $\mathbf{e} = \mathbf{x} - \mathbf{Ds}$. It is obviously that (10) is a convex programming and the solution can be obtained by some optimization software, such as the software cvx [10].

V. SIMULATION EXPERIMENT

In this section we present simulation results that demonstrate the performance of our method. For the remainder of this section we suppose that in the radar's beam scans process, the number of the responded impulse is 31; and the changes in beam scanning angle corresponding to the adjacent pulse interval is 0.1 degrees. The radar pattern is $p(\theta) = \text{sinc}^2(\theta_b^{-1}\theta)$, $\theta_b = 7.5°$. We did 1000 independent experiments count the error probability of the angle finding.

Fig. 2 to Fig. 4 demonstrate the performance of the four methods introduced in section IV, when the data extraction rates are one-half, one-quarter, and one-eighth respectively. The results of our basis pursuit algorithm are marked by pentacle and labeled 'BP'. While the ones of match pursuit, SCBM method, and the traditional maximum method are marked by diamond, rectangle and circularity, respectively. They are respectively labeled by 'MP', 'SCBM' and 'MAX'. The table 1 gives the corresponding mean square error of direction estimation by different methods.

The simulation results above show that the estimates based on sparse signal representation are better than the traditional maximum method and SCBM method. The proposed method can give a more accurate result even the volume of data is relatively few. Matching pursuit method can get the user's azimuth to the radar only by calculating the correlation between the atom and received signal. It is simple and easy to implement. However, due to influence of the correlation between adjacent atoms in the redundant dictionary, the performance of the match pursuit is inferior to standard basis pursuit method. The basis pursuit can optimize the optimal atom to get the solution, and the performance is much superior to others. However, compared with the matching pursuit method, the base basis pursuit method is more complex.

VI. CONCLUSION

Our results demonstrate that the angle finding can be significant improved if we incorporate the sparse information processing method into the radar antenna pattern modulated direction finding. The experimental results show that sparse signal representation based estimation is better than the traditional maximum method and the SCBM method. Although the base basis pursuit method is more computational complex in contrast to the matching pursuit method, the performance of the BP algorithm is much superior to the MP method for as much as the influence of the correlation between adjacent atoms in the redundant dictionary.

ACKNOWLEDGMENT

The authors thank anonymous reviewers for their valuable suggestions.

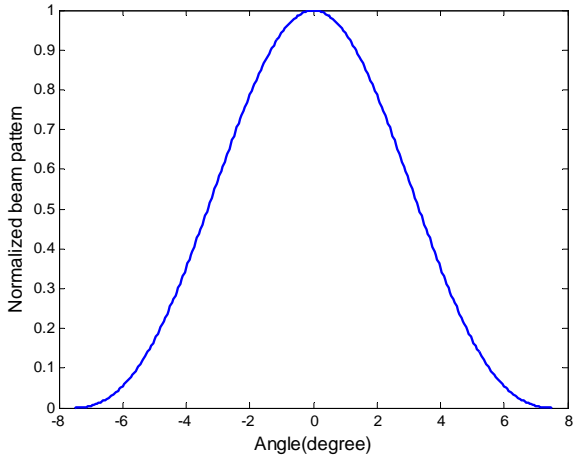

Fig. 1 the normalized beam pattern of radar antenna

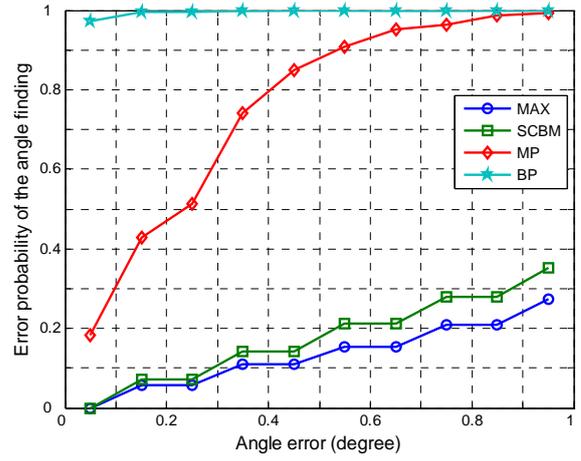

Fig. 3 Standard deviation of the direction error (extraction rate: one-quarter, SNR=5dB)

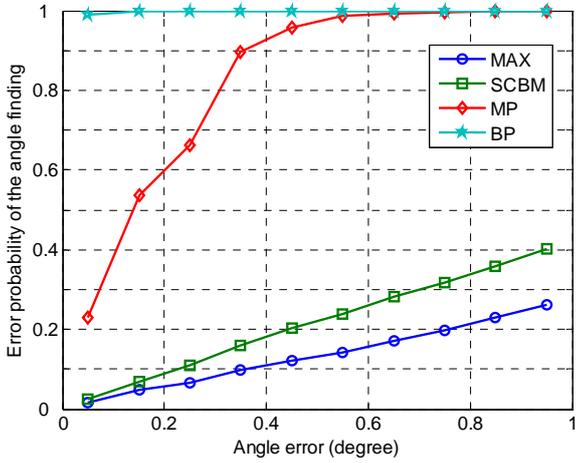

Fig. 2 Standard deviation of the direction error (extraction rate: one-half, SNR=5dB)

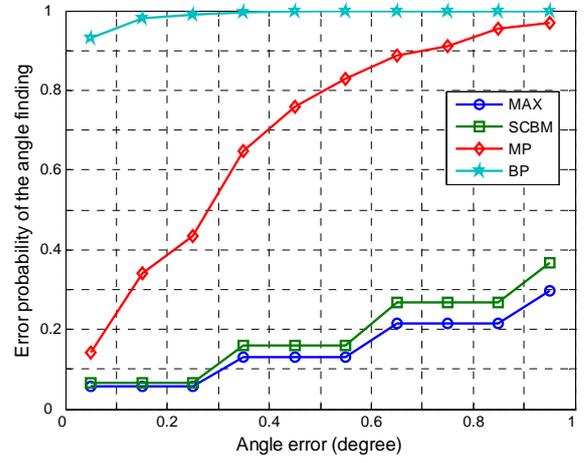

Fig. 4 Standard deviation of the direction error (extraction rate: one-eight, SNR=5dB)

TABLE I. MEAN SQUARE ERROR OF DIRECTION ESTIMATION (DEGREE)

| Sample's extraction rate | Maximum method | SCBM method | MP | BP |
| --- | --- | --- | --- | --- |
| One-half | 3.1101 | 2.2055 | 0.2498 | 0.0111 |
| One-quarter | 3.1777 | 2.9187 | 0.3666 | 0.0226 |
| One-eight | 3.0476 | 3.1662 | 0.5042 | 0.0741 |